\documentclass[aps, pra, twocolumn, showpacs, reprint]{revtex4-1}
\usepackage{graphicx,amsfonts}
\usepackage{epsf}
\usepackage{pstricks}
\usepackage{hyperref}
\begin{document}
\title{Superposing pure quantum states with partial prior information}
\author{Shruti Dogra}
\email{shrutidogra.iiserm@gmail.com \\ presently at Department of Physics, IIT Madras, Chennai, India.}
\affiliation{Optics and Quantum Information Group, 
The Institute of Mathematical Sciences, HBNI,  CIT Campus, Taramani, Chennai
600113, India}
\affiliation{Fakult\"{a}t Physik, Technische Universit\"{a}t Dortmund, D-44221 Dortmund,
Germany}
\author{George Thomas}
\email{georget@imsc.res.in}
\affiliation{Optics and Quantum Information Group, 
The Institute of Mathematical Sciences, HBNI,  CIT Campus, Taramani, Chennai
600113, India}
\author{Sibasish Ghosh}
\email{sibasish@imsc.res.in}
\affiliation{Optics and Quantum Information Group, 
The Institute of Mathematical Sciences, HBNI,  CIT Campus, Taramani, Chennai
600113, India}
\author{Dieter Suter}
\email{dieter.suter@tu-dortmund.de}
\affiliation{Fakult\"{a}t Physik, Technische Universit\"{a}t Dortmund, D-44221 Dortmund,
Germany}
\begin{abstract}
The principle of superposition is an intriguing feature of quantum mechanics, which 
is regularly exploited in many different circumstances. A recent work 
[PRL \textbf{116}, 110403 (2016)] shows that the fundamentals of quantum mechanics 
restrict the process of superimposing two unknown pure states, even though it is 
possible to superimpose two quantum states with partial prior knowledge. The prior 
knowledge imposes geometrical constraints on the choice of input states. We discuss 
an experimentally feasible protocol to superimpose multiple pure states of a 
$d$-dimensional quantum system and carry out an explicit experimental realization 
for two single-qubit pure states with partial prior information on a two-qubit NMR 
quantum information processor.
\end{abstract}
\pacs{03.67.Lx, 03.67.Ac}
\maketitle
 \section{Introduction \label{intro}} 
According to the postulates of  quantum theory, it is generally possible to generate 
superpositions of arbitrary pairs of pure states of a quantum system, unless there 
exists a superselection rule~\cite{dirac-book-1930,dass-qph-2013}.  However, a recent 
study showed that there exists no general quantum protocol for creating superpositions 
of a completely unknown pair of pure quantum states~\cite{alvarez-sr-2015,oszmaniec-prl-2016}. 
The difficulty of superimposing unknown quantum states was first discussed 
in Ref.~\cite{alvarez-sr-2015} in the context of quantum adders. 
Quantum states that are 
equivalent up to a global phase, represent the same physical state. 
Therefore the 
superposition of unknown quantum states that are equivalent, up to their global phases, 
may result in a relative phase between these states, and thus in different states. 
However, some partial prior knowledge about the states can be used to achieve the 
restricted type of superposition as suggested in a recent work~\cite{oszmaniec-prl-2016}. 
As shown in Ref.~\cite{oszmaniec-prl-2016}, two unknown quantum states 
 $\vert \psi_1 \rangle$ and $\vert \psi_2 \rangle$ can be 
superposed, if their overlaps with a  reference state $\vert \chi \rangle$ are known 
and nonzero. For the superposition of two $d$-dimensional states, a tripartite system 
of dimension $2d^2$ is used. The corresponding state is initialized into 
$(a \vert 0 \rangle + b \vert 1 \rangle) \vert \psi_1 \rangle \vert \psi_2 \rangle$,
with arbitrary complex coefficients $a$, $b$.
This state is subsequently transformed by a three-party controlled-SWAP gate. Finally, 
two projection operators are constructed using the reference state $\vert \chi \rangle$
and its overlaps ($\vert \langle \chi \vert \psi_i \rangle \vert$) with the states to 
be superimposed. The application of these projectors generates a state proportional 
to $(a \kappa_2 \vert \psi_1 \rangle + b \kappa_1 \vert \psi_2 \rangle)$,
where $\kappa_i= \langle \chi \vert \psi_i \rangle/\vert \langle \chi \vert \psi_i \rangle \vert$.
\par
In general, for the sake of quantum computation, it may be useful to experimentally 
superpose unknown quantum states~\cite{lamanta-qph-2017}. For the past few decades, there 
has been a growing interest for more feasible, robust experimental quantum 
computation models~\cite{nielsen-book-02, suter-book-2004, suter-book-2008,ladd-nature-2010}.
Experimental realization of superposition of unknown quantum states is significant, 
not only as a quantum computational task, but also as a fundamental principle. There 
exist experimental techniques based on photons~\cite{hu-pra-2016}, nuclear 
spins~\cite{li-pra-2016}, and super conducting qubits~\cite{unai-qph-2016} that 
implemented the superposition protocol discussed in~\cite{oszmaniec-prl-2016}. 
In Ref.~\cite{hu-pra-2016}, the superposition of two photonic states is realized. 
The controlled-SWAP implementation was a challenge here; therefore, an effective 
controlled-SWAP operation was implemented which includes post-selection and 
is a non-unitary operation. Another work~\cite{li-pra-2016} presents the experimental 
implementation of the superposition protocol~\cite{oszmaniec-prl-2016} using three 
nuclear spins, where the controlled-SWAP gate was implemented via numerically 
optimized pulses. This was followed by a three-qubit tomography and, subsequently, 
tracing out first and third qubits numerically to imitate projective measurements. 
A transmons-based implementation of Refs.~\cite{alvarez-sr-2015, oszmaniec-prl-2016} 
was realized on the IBM Quantum Experience~\cite{unai-qph-2016}. This scheme implemented 
an optimal quantum circuit obtained using genetic algorithm techniques, but its 
operation is limited to specific input states. 

\par 
The present work experimentally realizes a full protocol to perform the desired 
superpositions of pure states of a quantum system, addressing all the aspects 
discussed in Ref.~\cite{oszmaniec-prl-2016}.
The experiment friendly superposition protocol discussed here overcomes the 
experimental inefficiencies reported in Ref.~\cite{li-pra-2016}.
Moreover, this is a two-qubit 
based experimental implementation to superpose two single-qubit states contrary 
to the existing implementation that used three physical qubits~\cite{li-pra-2016}.
The protocol is further generalized to superpose $n$ higher-dimensional quantum states.
A detailed comparison between our experimentally implemented
protocol with that of existing experimental implementations in terms of the 
success probabilities is carried out.
We also analyze the  enhancement in the success probabilities
associated with the desired superpositions for different prior information.

\par
The material in this paper is arranged as: 
theoretical development of the experiment-friendly superposition protocol
is described in Section~\ref{theory}.
Further, experimental implementation using a system of two-nuclear spins 
is given in Section~\ref{experiment}.
The extension of our scheme to superpose $n$ higher-dimensional quantum states is 
discussed in Section\ref{nqudits}. The comparison of the success probabilities
with respect to previously implemented superposition protocol~\cite{li-pra-2016},
and its enhancement subject to prior information is discussed in
section~\ref{discussion}. This is followed by the concluding section~\ref{conclusion}.
\section{Theoretical scheme \label{theory}} 
Let us consider the superposition of two arbitrary states $|\Psi_1\rangle$,
and $|\Psi_2\rangle$, with desired weights of superposition ($a$ and $b$), and
whose respective inner products $\langle \chi|\Psi_i \rangle$ with a known
referential state $|\chi\rangle$ are given.
It is well known that a state $|\Psi\rangle$ and $e^{\iota \gamma}|\Psi\rangle$
represent the same physical states, despite different values of the overall phase `$\gamma$'.
However the superposition of these states depend upon the values 
of the respective overall phases of the constituent states.
While the global phase of a state is intangible, 
it is possible to determine the overall phase of a state  with respect to a reference 
state. Here we use the partial prior information given in terms of the inner products $\langle \chi|\Psi_i \rangle$
 to obtain the overall phase factors, 
$e^{\iota \gamma}=\langle \chi|\Psi_i \rangle/|\langle \chi|\Psi_i \rangle|$.
The details of the protocol are worked out in the following stanzas. 
Thus, for the class of states $|\Psi_i\rangle=e^{\iota \gamma_i}|\psi_i\rangle$, that are 
equivalent to each other upto an overall phase, $\gamma_i \in [0,2\pi]$, the 
desired superimposed state may be written as, $a |\psi_1 \rangle + b |\psi_2\rangle$.

\par
Beginning with an explicit analysis for the superposition of two single-qubit pure 
states, we consider a system of two coupled spin-$1/2$ particles (denoted here as 
A and X) under the action of a Hamiltonian
\begin{equation}
 H= -\Omega_A A_{z}\otimes \mathbb{I}_{X} -\Omega_X \mathbb{I}_{A} 
 \otimes X_{z} + J A_{z} \otimes X_{z}, \label{ham}
\end{equation}
where $\Omega_A$ ($\Omega_X$) is the resonance frequency and
$A_{z}$ ($X_{z}$) is the $z$-component of angular momentum for spin $A$ ($X$).
 $J$ represents the scalar coupling constant. 
$\vert 0 \rangle_A, \vert 1 \rangle_A$ ($\vert 0 \rangle_X, 
\vert 1 \rangle_X$) are the eigenvectors of $A_{z}$ ($X_{z}$) with
eigenvalues $+1/2, -1/2$ respectively. The single-qubit pure states of our system 
are encoded in the eigenbasis $\{ \vert 00 \rangle, \vert 01 \rangle, 
\vert 10 \rangle, \vert 11 \rangle\}$ of the Hamiltonian $H$. 
We use the subspace spanned by $\vert 0 0 \rangle$, $\vert 0 1 \rangle$ of $H$ to 
store the single-qubit input state  $\vert \Psi_1 \rangle = c_{0 0} \vert 0 \rangle 
+ c_{0 1} \vert 1 \rangle$, where $\vert c_{00}\vert^2+\vert c_{01}\vert^2=1$, while 
the subspace spanned by the two remaining levels is used to store the input state 
vector $\vert {\Psi}_2 \rangle = c_{1 0} \vert 0 \rangle +  c_{1 1} \vert 1 \rangle$,
where $\vert c_{10}\vert^2+\vert c_{11}\vert^2=1$. The state of the two-qubit system 
($A+X$) is then
\begin{equation}
\vert \Psi \rangle^{'} = a \vert 0 \rangle \otimes e^{\iota \gamma_1} \vert {\psi}_1 \rangle + b \vert 1 \rangle \otimes e^{\iota \gamma_2} \vert {\psi}_2 \rangle; 
\quad \vert a \vert^{2} + \vert b \vert^{2}=1,
\label{coded}
\end{equation}
where $a$ and $b$ are the weights of the superposition. In Eq.~(\ref{coded}), the 
first qubit is the ancilla and the second qubit is the system-qubit. 
The superposition protocol that 
we propose here generates the desired superimposed state,
irrespective of the values of phase factors (say $e^{\iota \gamma_j}$ with $j^{th}$ 
input state)\cite{oszmaniec-prl-2016}. 
Given any fixed state $\vert \chi \rangle$ of the system qubit (such that 
$\langle \chi \vert {\psi}_i \rangle \neq 0$),
prior knowledge of the inner products 
$\langle \chi \vert {\psi}_1 \rangle$ and $\langle \chi \vert {\psi}_2 \rangle$ is exploited 
to find the phases $e^{\iota \gamma_j}$.
Using this information, we construct a phase gate ($e^{\iota \theta_{z} (A_{z} 
\otimes \mathbb{I}_{X})}$), that implements a $z-$rotation on the first qubit by an angle 
$\theta_{z}=\frac{\gamma_1-\gamma_2}{2}$, leading to the state,
\begin{equation}
\vert \Psi \rangle^{''} \equiv   e^{\iota \frac{\gamma_1+\gamma_2}{2}}
(a \vert 0 \rangle \vert \psi_1 \rangle + b \vert 1 \rangle \vert \psi_2 \rangle). \label{eq3}
 \end{equation}
Thus the phases with the individual single-qubit states are modified, and appear 
as an overall phase of the two-qubit state. In Appendix~\ref{appA} a detailed view
of an alternative protocol is given to encode the states 
$\vert \psi_1 \rangle$, $\vert \psi_2 \rangle$ and to get rid of their phases 
$e^{\iota \gamma_1}$, $e^{\iota \gamma_2}$ respectively. Further, a Hadamard gate 
on the first-qubit in Eq.~(\ref{eq3}) leads to the state (ignoring the overall 
phase $e^{\iota \frac{\gamma_1 + \gamma_2}{2}}$), 
\begin{equation}
 \vert \Psi \rangle^{'''} \equiv \frac{\vert 0 \rangle}{\sqrt{2}} 
(a \vert \psi_1 \rangle +  b \vert \psi_2 \rangle)  + \frac{\vert 1 \rangle}{\sqrt{2}}
(a \vert \psi_1 \rangle -  b \vert \psi_2 \rangle). \label{eq6} 
\end{equation}
Depending upon the state of the first qubit, one can choose between the sum or 
difference of the single-qubit states $\vert \psi_1 \rangle$ and $\vert \psi_2 
\rangle$: a measurement on the first qubit in the basis $\{ \vert 0 \rangle , 
\vert 1 \rangle\}$ gives rise to the state, $a \vert \psi_1 \rangle +  b \vert 
\psi_2 \rangle$ of the second qubit (in case of outcome $\vert 0 \rangle$) which 
is proportional to the desired superposed state, $N_{\psi}(a \vert \psi_1 \rangle 
+ b \vert \psi_2 \rangle)$ ($N_{\psi}$ being the normalization constant), obtained 
with a success probability $N_{\psi}^2/2$. Thus, with the help of only one 
ancillary qubit, we are able to superpose two single-qubit states.
Also, `$e^{\iota \gamma_i}$' does not show up in the final superposed
state, which implies that the overall phase factors of the constituent states
do not alter the resultant superimposed state in this protocol.

\par
In the present context, no-go theorems concerning the implementation of
unknown quantum operations~\cite{thompson-njp-2018, arjau-njp-2014,nicolai-pra-2014} 
are circumvented by using the general protocol, that creates ``arbitrary" pairs of 
input states within the given constraints. It is important to note that no extra 
information regarding arbitrary pairs of input states is used further in the 
superposition protocol.
\section{Experimental implementation \label{experiment}}
\begin{figure}
 \centering
 \includegraphics[scale=1]{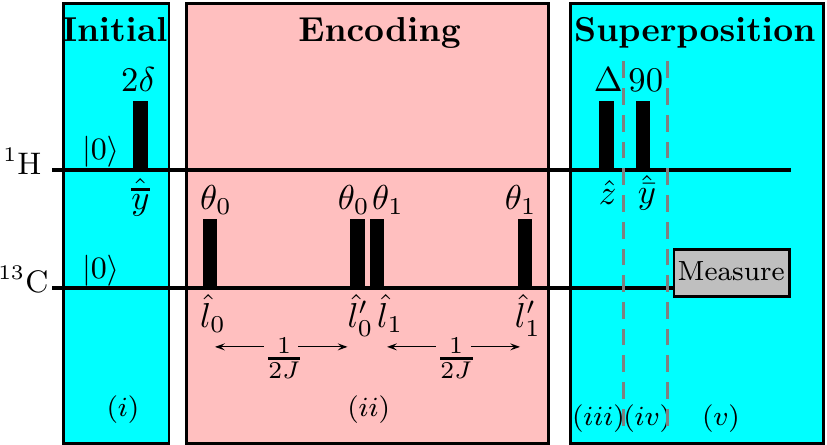}
 \caption{(Colour online) NMR pulse sequence to obtain a superposition of two single-qubit states 
 starting with the pseudo-pure state $\vert 00 \rangle$. The two channels show the 
 operations on ancilla ($^{1}\rm{H}$) and system qubits ($^{13}\rm{C}$) respectively. 
Pulse sequence is divided into three parts, shown as separate
 blocks of different colors.
 Also, various steps are numbered from $(i)$-$(v)$. The radio-frequency pulses are shown as rectangles, 
 with respective angles of rotations mentioned at the top and the axes of rotations 
 specified at the bottom. The arbitrary  rotation axes are 
 $\hat{l}_0=\cos(\frac{3\pi}{2}+\phi_0)\hat{x}+\sin(\frac{3\pi}{2}+\phi_0)\hat{y}$,
 $\hat{l}_0'=\cos(\phi_0)\hat{x}+\sin(\phi_0)\hat{y}$,
 $\hat{l}_1=\cos(\pi+\phi_1)\hat{x}+\sin(\pi+\phi_1)\hat{y}$, and 
 $\hat{l}_1'=\cos(\frac{\pi}{2}+\phi_1)\hat{x}+\sin(\frac{\pi}{2}+\phi_1)\hat{y}$.
At the end of the sequence, a single-qubit measurement is performed on the system qubit. 
\label{pp}}
\end{figure}

The NMR pulse sequence to carry out weighted superposition of two single-qubit 
states is shown in Fig.~\ref{pp}, 
where the first channel corresponds to the 
ancillary-qubit $A$ and the second channel corresponds to the system qubit 
$X$ (here labeled as $^{1}\rm{H}$ and $^{13}\rm{C}$ respectively).
Pulse sequence is divided into three
blocks: initial, encoding and superposition as mentioned in 
Fig.~\ref{pp}. In the first block, system and ancillary 
qubits are jointly initialized 
in state $\vert 00 \rangle$. A single-qubit rotation by an angle $2\delta$ about 
the $\hat{\overline{y}}-$axis 
is applied on the ancillary qubit, generating the state 
$a \vert 00 \rangle +  b \vert 10 \rangle$ (with $a=\cos \delta$ and $b=\sin \delta$). 
Second block, labeled as `encoding', encodes the 
arbitrary pair of single qubit states.
This is achieved by two two-qubit controlled operations, that encode second qubit 
with state $\vert \psi_1 \rangle$, when first qubit is in state $\vert 0 \rangle$ 
and with state $\vert \psi_2 \rangle$ when first qubit is in state $\vert 1 
\rangle$. Each controlled-operation is achieved by a controlled-rotation of 
second-qubit by an angle $(\theta_j)_{n_j}$ where state of the first qubit, 
$\vert j \rangle$ ($j \in \{ 0,1 \}$) is the control. The axis of rotation, 
$\hat{n}_j=\cos (\phi_j) \hat{y} + \sin (\phi_j)\hat{x}$. 
At the end of this step (labeled as $(ii)$ in Fig.~\ref{pp}), 
joint state of system and ancilla is
given by $a \vert 0 \rangle |\psi_1 \rangle +  b \vert 1 \rangle |\psi_2 \rangle$,
such that the encoded state $|\psi_j \rangle$ is parametrized by 
$\{\theta_{j-1}, \phi_{j-1}\}$ ($j=1,2$).
This encoded two-qubit state is then fed into the block named
`superposition', wherein
possible overall phases of the arbitrary input states
$\psi_1$ and $\psi_2$ are taken care of by applying a 
$z-$pulse of angle $\Delta=\frac{\gamma_1-\gamma_2}{2}$ on the first qubit, leading
to the state given in Eq.~(\ref{eq3}).
This 
is followed by a pseudo-Hadamard gate on the ancillary qubit, which
is a $90^0$ pulse about $-y$ direction, leading to the joint state of system and ancilla 
as given in Eq.~(\ref{eq6}). 
A partial read out of 
the system qubit leads to the expected superposed state. In all the experiments, 
the referential state ($\vert \chi \rangle$) is chosen as $\vert 0 \rangle$. 
\par As discussed in the theoretical scheme, the measurement consists of a projective 
measurement on the first qubit ($\vert 0 \rangle \langle 0 \vert \otimes \mathbb{I}_{2X2}$),
followed by a partial-trace operation that retains the state of the second qubit. The 
measurement applies therefore \textit{only} to the subspace spanned by the eigenvectors 
$\vert 00 \rangle$ and  $\vert 01 \rangle$ of $H$. Experimentally, the corresponding 
information is contained in the coherence between these two states. Thus the final 
superposed state is recovered from a two-dimensional subspace by partial quantum 
state tomography. This approach may also be useful in different experiments as a 
replacement of projective readout. The desired single-qubit density operator is obtained 
by a set of two operations: (i) direct readout, to obtain the 
information about the single-quantum coherence between states $\vert 00 \rangle-\vert 01 \rangle$ and 
(ii) application of a gradient ($G_z$), followed by a $90^0$ pulse about $y-$axis ($(\frac{\pi}{2})_y^2$)
on the second qubit, to obtain the relative populations of the energy levels $\vert 00 \rangle$ and $\vert 01 \rangle$. 
In both cases, we observe the spectral line corresponding to transition $\vert 00 
\rangle-\vert 01 \rangle$. The resultant single-qubit density operator is un-normalized 
in this protocol. The normalization constant for the desired part of the density operator 
can be obtained experimentally by measuring the  sum of the populations of states 
$\vert 00 \rangle$ and $\vert 01 \rangle$. This is achieved by applying a gradient to 
dephase the coherences, followed by a spin-selective $90^0$ pulse on the first qubit 
($G_z (\frac{\pi}{2})_y^1$). A readout of the resultant NMR spectrum of the first qubit 
provides the normalization constant for the desired subspace. This normalization factor 
is then used to completely characterize the final state density operator of the superposed 
state.
\par The pulse sequence shown in Fig.~\ref{pp} is implemented experimentally on a sample 
consisting of $^{13}\rm{C}$ labeled Chloroform in deutrated Acetone. The experiments were 
performed on a 500 MHz Bruker Avance II NMR spectrometer  with a QXI probehead. 
All  pulses were high power, short duration RF pulses applied to the $^{1}\rm{H}$ and 
$^{13}\rm{C}$ spins on resonance. Scalar coupling constant, $J=215$ Hz. The spin-spin relaxation times ($T_2^{*}$) of the  $^{1}\rm{H}$ 
and $^{13}\rm{C}$ spins were 540 ms and 170 ms, respectively. Nuclear spin systems at thermal 
equilibrium are in a mixed state. The system was thus initialized into a pseudo-pure 
state, $\vert 00 \rangle$  by spatial averaging~\cite{cory-pd-1998} with a fidelity of 
$0.999$. Starting from this pseudo-pure state, various pairs of single-qubit states 
($\vert \psi_1 \rangle$ and $\vert \psi_2 \rangle$) were encoded on a two-qubit system, 
as described earlier. 

\par
\begin{widetext}
\begin{figure*}
 \centering
 \includegraphics[scale=1]{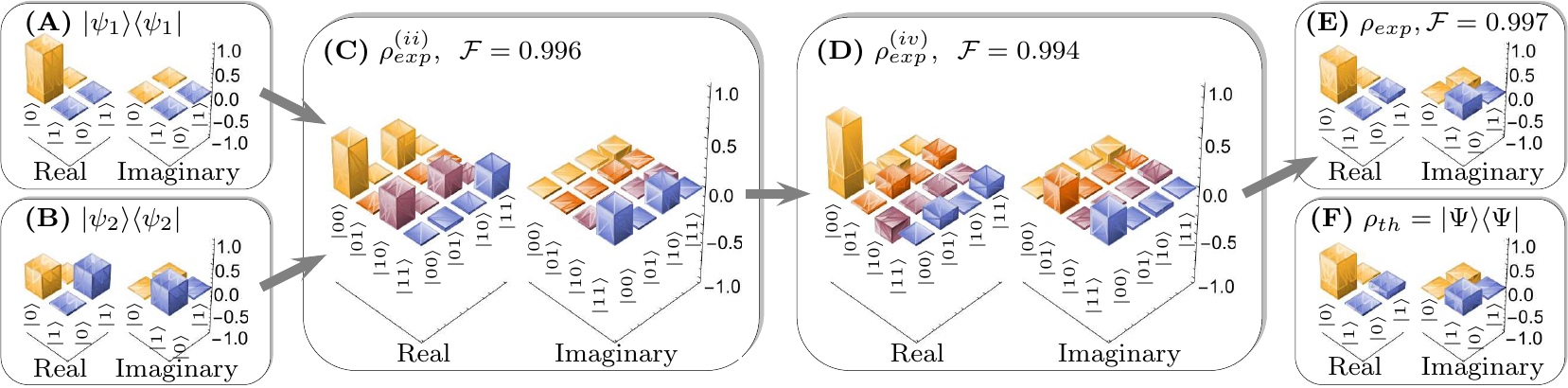}
 \caption{(Colour online) (A) and (B) show the theoretical input states from 
 dataset 3 of Table~\ref{table1}, part (C) contains the two-qubit state after 
 encoding ($\rho_{exp}^{(ii)}$), (D) represents the state obtained at the 
 end of step ($iv$) of the pulse sequence ($\rho_{exp}^{(iv)}$), parts (E) 
 and (F) show the final experimentally obtained ($\rho_{exp}$) corresponding to
 step $(v)$ and theoretically 
 expected ($\rho_{th}$) single-qubit superposed states respectively. \label{tomo}}
\end{figure*}
\end{widetext}

In order to ensure the accuracy of this experimental implementation, two-qubit 
density operators were tomographed at the end of step $(ii)$ and $(iv)$ of the 
pulse sequence (Fig.~\ref{pp}), thus obtaining the state after encoding 
($\rho_{exp}^{(ii)}$) and the state before the measurement ($\rho_{exp}^{(iv)}$) 
respectively. The two-qubit states were completely reconstructed with a set 
of four operations: $\{\mathbb{II}, \mathbb{IX}, \mathbb{IY}, \mathbb{XX} \}$, 
where $\mathbb{X(Y)}$ refers to spin-selective $90^0$ pulse along $x(y)$-axis. 
Single-qubit density operator of the system qubit 
is obtained through two operations on the system qubit:
$\{ \mathbb{I}, \,G_z \mathbb{Y} \}$, where $G_z$ is the 
non-unitary gradient implementation about $z-$axis. The resultant 
single-qubit reduced density operator is then normalized as described earlier
in this section.
The fidelity between the theoretically expected ($\rho_{t}$) and the experimentally 
obtained ($\rho_{e}$) states were measured using the following expression,
\begin{equation}
\mathcal{F} = Tr(\rho_{e}  \rho_{t})/\sqrt{Tr(\rho_{e}^{2}) Tr(\rho_{t}^{2})}.
\end{equation}
\begin{center}
\begin{table}
\caption{Summary of experimental results. 
\label{table1}}
\begin{center}
\begin{tabular}{lllll}
\hline
\hline
S.No.& Input state $\vert \psi_1 \rangle$ & Input state $\vert \psi_2 \rangle$ & $\frac{a}{b}$ 
& $\quad\mathcal{F}$ \\
\hline 
1 & $\vert 0 \rangle$ & $\frac{1}{\sqrt{2}}(\vert 0 \rangle + \vert 1 \rangle)$ & $1$ 
& 0.996 \\
 2 & $\vert 0 \rangle$  & 
 $\frac{1}{\sqrt{2}}(\vert 0 \rangle + e^{\frac{\iota \pi}{4}}\vert 1 \rangle)$ & $1$ 
& 0.995 \\
 3 & $\vert 0 \rangle$ & $\frac{1}{\sqrt{2}}(\vert 0 \rangle + e^{\frac{\iota \pi}{2}}\vert 1 \rangle)$ & $1$ 
& 0.997 \\
 4 & $\vert 0 \rangle$ & $\frac{1}{\sqrt{2}}(\vert 0 \rangle + e^{\iota \pi}\vert 1 \rangle)$ & $1$ 
& 0.997 \\
 5 & $\frac{1}{2}(\vert 0 \rangle + \sqrt{3}\vert 1 \rangle)$ 
& $\frac{1}{2}(\sqrt{3} \vert 0 \rangle + \vert 1 \rangle)$ & $1$ 
& 0.998 \\
 6 & $\frac{1}{2}(\vert 0 \rangle + e^{\frac{\iota \pi}{4}} \sqrt{3}\vert 1 \rangle)$ 
& $ \frac{1}{2}(\sqrt{3} \vert 0 \rangle + e^{\frac{\iota 2\pi}{3}} \vert 1 \rangle)$ & $1$ 
& 0.974 \\
 7 & $\frac{1}{2}(\vert 0 \rangle + \sqrt{3}\vert 1 \rangle)$ 
& $\frac{1}{2}(\sqrt{3} \vert 0 \rangle + \vert 1 \rangle)$ & $2$ 
& 0.999 \\
 8 & $\frac{1}{2}(\vert 0 \rangle + \sqrt{3}\vert 1 \rangle)$ 
& $\frac{1}{2}(\sqrt{3} \vert 0 \rangle + \vert 1 \rangle)$ & $3$ 
& 0.999 \\
 9 & $\frac{1}{2}(\vert 0 \rangle + \sqrt{3}\vert 1 \rangle)$ 
& $\frac{e^{\frac{2\pi\iota}{3}}}{2}(\sqrt{3} \vert 0 \rangle + \vert 1 \rangle)$ & $1$
& 0.999 \\
 10 & $\frac{1}{2}(\vert 0 \rangle + e^{\frac{\iota \pi}{4}} \sqrt{3}\vert 1 \rangle)$ 
& $ \frac{e^{\frac{2\pi\iota}{3}}}{2}(\sqrt{3} \vert 0 \rangle + e^{\frac{\iota 2\pi}{3}} \vert 1 \rangle)$ & $1$ 
& 0.981 \\
 11 & $ \vert 0 \rangle$ 
& $ \sin \frac{\pi}{36} \vert 0 \rangle + \cos \frac{\pi}{36} \vert 1 \rangle$ & $1$
& 0.988 \\
\hline 
\hline
\end{tabular} 
\end{center}
\end{table}
\end{center}
Table~\ref{table1} summarizes the results of various experiments, with columns 2 
and 3 showing the single-qubit pure states to be superposed, and column 5 contains 
the fidelity ($\mathcal{F}$) between the experimentally superposed states and the 
theoretically expected ones. In the datasets numbered 1-4 of Table~\ref{table1}, 
we have, $\vert \psi_1 \rangle=\vert 0 \rangle$, and $\vert \psi_2 \rangle
=\frac{1}{\sqrt{2}}(\vert 0 \rangle + e^{\iota \phi_2} \vert 1 \rangle)$, with 
$\phi_2 \in \{0, \frac{\pi}{4}, \frac{\pi}{2},\pi \}$. Each of these pairs 
corresponds to the same two conical sections as per their Bloch sphere 
representations. Similarly, the datasets numbered 5 and 6 of the Table~\ref{table1} 
show the superposition between two pairs of states from the same respective 
conical sections. A detailed tomographic analysis corresponding to  dataset 3
(Table~\ref{table1}) is shown in Fig.~\ref{tomo}. We also generated  superpositions 
of the same constituent states with different weights, as given in datasets 
5, 7 and 8 of Table~\ref{table1}. For completeness, the experiments were performed 
with different overall phases of the input states. These phase factors were 
introduced while encoding the states $\vert \psi_1 \rangle$ and $\vert \psi_2 
\rangle$, by applying a pulse of angle $2\delta$ about the axis $`\hat{l}'$ which is 
aligned with $y-$axis at an angle $\pi+\gamma_2$ (Fig.~\ref{pp}). The encoded state 
is  thus of the form, $a \vert 0 \rangle \vert \psi_1 \rangle + e^{\iota \gamma_2} 
b \vert 1 \rangle \vert \psi_2 \rangle $. Experiments were performed for two pairs 
of states shown in datasets  9 and 10 in Table~\ref{table1}. In both cases, 
$\gamma_2=120^0$ and the remaining parameters were same as those of sets 5 and 6 
in Table~\ref{table1}. Now compare the datasets 5 with 9 and 6 with 10. As expected, 
the presence or absence of the overall phase does not affect the final superposed 
state. The efficacy of this experimental scheme does not directly depend upon the 
values of the overlaps ($|\langle \chi \vert \psi_i \rangle|$). This is evidenced 
by the dataset 11 of Table~\ref{table1}, where $\vert \psi_2 \rangle$ is very close 
to $\vert \chi^{\bot} \rangle$ (orthogonal to $\vert \chi \rangle$). Table~\ref{table1} 
shows that even if we choose the pair of input states ($\vert \psi_1 \rangle, 
\vert \psi_2 \rangle$) outside the set $\{(\vert \psi_1 \rangle, 
\vert \psi_2 \rangle): \vert \langle \chi \vert \psi_1 \rangle \vert=\textrm{constant,}\; 
\vert \langle \chi \vert \psi_2 \rangle \vert=\textrm{constant}\}$, our procedure 
still generates the expected superposition state $a \vert \psi_1 \rangle +b \vert 
\psi_2 \rangle$ with  high accuracy.
\section{Superposition of multiple qudits \label{nqudits}}
Our procedure can be readily extended to the superposition of arbitrary pure 
states of $n$  qudits ($d$-dimensional quantum system)~\cite{oszmaniec-prl-2016}. 
Let $a_1,\, a_2,\dots a_{n}$ be the desired coefficients for creating a
superposition of $n$ ($d$-dimensional) states $\vert \Psi_1 \rangle_{d}$, 
$\vert \Psi_2 \rangle_{d}$, $\dots$ $\vert \Psi_{n} \rangle_{d}$.
This requires a hybrid $n\times d-$dimensional qu$n$it-qudit system, where 
the qu$n$it ($n-$dimensional quantum system) acts as an ancilla (as before) 
and the qudit acts as the system. 
For simplicity, we use a vector representative $|\Psi\rangle_j$ 
to represent the set of states $e^{\iota \gamma_j}|\Psi\rangle_j$, 
where $\gamma_j \in [0,2\pi]$. Consider now a $d$-dimensional referential 
state $\vert \chi \rangle_{d}$, whose non-zero overlaps, $\vert \langle 
\chi \vert \Psi_j \rangle_{d} \vert^2=c_j$, ($j \in \{1,2,\dots n \}$) are known. 
Following the same protocol as before, every qudit state is encoded in the 
$n \times d$  basis vectors of the hybrid qu$n$it-qudit system: $\vert j0 
\rangle, \, \vert j1 \rangle, \, \vert j2 \rangle, \dots \vert j(d-1) 
\rangle$ where $j \in \{ 0,1,\dots n-1\}$. The phases of the constituent 
states are taken care of by using the information of overlaps of respective 
constituent states with the referential state (see Appendix~\ref{appA}). This is then 
followed by Fourier transformation of the qu$n$it, which is in fact the 
generalization of the Hadamard operation to higher-dimensional states
\cite{dogra-ijqi-2015}. The resultant state, which is a generalization of 
the two-qubit state in Eq.~(\ref{eq6}), is 
\begin{equation}
\frac{1}{N\sqrt{n}} \sum_{j=0}^{n-1} \left(
 \vert j \rangle_{n} \otimes \sum_{k=1}^{n} \left(
f^{j(k-1)} a_{k} \vert \Psi_{k} \rangle_{d}
\right) \right),
 \label{eq10}
\end{equation}
where $f= e^{\iota \frac{2\pi}{n}}$, is the $n^{th}$ root of unity and $N$ 
is the normalization constant. An arbitrary superposition of $n$ pure states 
of a qudit is then obtained by the projective measurement 
$\vert 0 \rangle_{n} \langle 0 \vert_{n} \otimes \mathbb{I}_{d \times d}$
subsequently tracing out the qu$n$it. The final state is 
a superposition of $n$ $d$-dimensional states, which along with the 
information of overall phase factors of the constituent ($n$-qudits) 
states is (from Appendix~\ref{appA}),
\begin{equation}
 \vert \Psi \rangle = \frac{N_{\Psi}}{N \sqrt{n}} \sum_{k=1}^{n} a_k 
 \left( \prod_{(j\neq k, j=1)}^{n}{\frac{\langle \chi \vert \Psi_j 
 \rangle_{d}}{\sqrt{c_j}}} \right) \vert \Psi_{k} \rangle_{d}, \label{eq16}
\end{equation}
where $N_{\Psi}$ is a constant that normalizes the un-normalized state 
obtained after the projective measurement.  The superposed state $\vert 
\Psi \rangle$ (Eq.~(\ref{eq16})) is obtained with the success probability,
\begin{equation}
 P=\frac{N_{\Psi}^2}{N^2 n}=\frac{ \prod_{j=1}^{n}c_j}
 {\sum_{j=1}^{n}  a_j^2 c_j}  \frac{ N_{\Psi}^2}{n}. \label{prob}
\end{equation}
\section{Discussion \label{discussion}}
As per superposition protocol discussed in Ref.~\cite{oszmaniec-prl-2016},
a projector $| \mu \rangle \langle \mu |$ 
(where $|\mu\rangle \propto \sqrt{c_1}|0\rangle + \sqrt{c_2}|1\rangle$) is applied on first qubit to obtain
the superposed state. It is discussed in~\cite{li-pra-2016}, that
precision of the implementation of this operator highly depends upon the values of
$|\langle \chi|\psi_1 \rangle |$ and $|\langle \chi|\psi_2 \rangle |$. 
Smaller values of these overlaps
lead to huge errors. Detailed analysis of this issue is carried 
out by Li et al (\cite{li-pra-2016}), where it is shown that when any of 
the overlap values ($|\langle \chi|\psi_1 \rangle |$, $|\langle \chi|\psi_2 \rangle |$)
approaches zero, the protocol unexpectedly results the final
states with poor fidelities. It has been clearly stated in Ref.~\cite{li-pra-2016}
that the malfunctioning of the protocol, as
$|\langle \chi|\psi_1 \rangle |$ or $|\langle \chi|\psi_2 \rangle | \rightarrow 0$,
is mainly due to experimentally unavoidable imprecisions in
the implementation of $| \mu \rangle \langle \mu | \otimes I \otimes I$
projection operator.
However in the protocol implemented here, no such projector is used. Instead, we
implement a Hadamard operator which due to its ease to implement,
neatly gives the resultant
state. This is also reflected in one of our experimental results
(Table~\ref{table1}, dataset no. 11) where, despite very small value
of the overlap between the referential state and the constituent state, 
experimental superimposed state is obtained with good fidelity. 
Thus the precision of our protocol is actually 
independent of the values of these overlaps, which 
makes this protocol more experimentally feasible.
\\
A more close analysis of success probabilities 
obtained in different superposition protocols, and for 
different amount of prior information is given in following
sub-sections.
\subsection{Comparison between general two-qubit and three-qubit based implementations \label{appB}}
\begin{figure}[h!]
 \centering
 \includegraphics[scale=1]{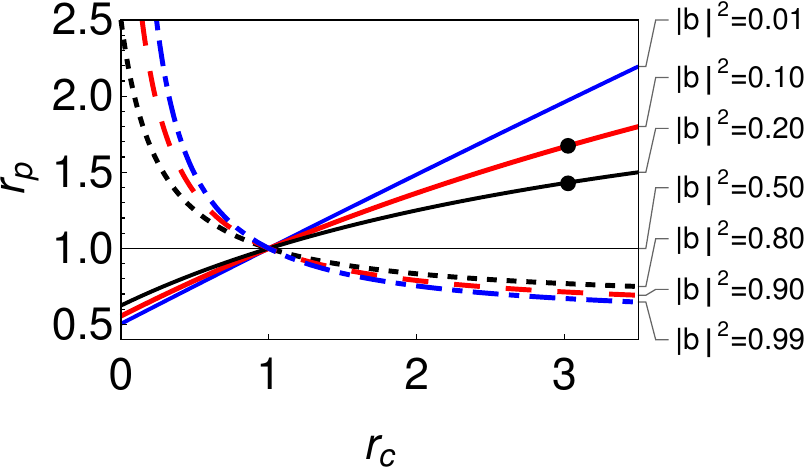}
 \caption{(Colour online) The variation of $r_p=P_2/P_3$ is shown with the ratio 
 of overlaps, $r_c=c_2/c_1$ corresponding to different values of
 $\vert b \vert^2$. Different curves correspond to different 
 values of $\vert b \vert^2$,
 that are specified on the right side of the plot. Two black points
 on the curves for $\vert b \vert^2=0.1, \, 0.2$ correspond to 
 experimental conditions of the datasets numbered $7$ and $8$ of Table~\ref{table1}.
 \label{plot}}
\end{figure}

In this section, we compare the success probabilities 
obtained in our scheme with that of
previously implemented scheme~\cite{oszmaniec-prl-2016, li-pra-2016} to carry out the 
superposition of two single-qubit states. 
With the purpose of comparison, we start with same amount 
of resources. Thus we use the protocol discussed in Section~\ref{theory} to 
obtain the present two-qubit based scheme 
from the existing three-qubit based scheme~\cite{oszmaniec-prl-2016}
to superimpose two single-qubit pure states. Recalling Eq.\ref{eq12n},
the resultant superposed state is given as,
\begin{equation}
  \sqrt{\frac{c_1 c_2}{2(c_1\vert a \vert^2+c_2\vert b \vert^2)}} \left( a 
  \frac{\langle \chi \vert \psi_2 \rangle}{\vert \langle \chi \vert \psi_2 \rangle \vert}
  \vert \psi_1 \rangle + b 
  \frac{\langle \chi \vert \psi_1 \rangle}{\vert \langle \chi \vert \psi_1 \rangle \vert}
  \vert \psi_2 \rangle \right) \label{eq13}.
\end{equation}
The success probability in this case is given as $P_{2}=\frac{c_1 c_2}{2(c_1\vert a \vert^2+c_2\vert b \vert^2)}N_{\psi}^2$.
Here $N_{\psi}$ is the normalization factor for state 
$a \vert \psi_1 \rangle + b \vert \psi_2 \rangle$ (where $\sqrt{\vert a \vert^2+\vert b \vert^2}=1$). 
Recalling the treatment in a three-qubit based protocol~\cite{oszmaniec-prl-2016, li-pra-2016}, 
the resultant state in that case is given as,
\begin{equation}
  \sqrt{\frac{c_1 c_2}{c_1+c_2}} \left( a 
  \frac{\langle \chi \vert \psi_2 \rangle}{\vert \langle \chi \vert \psi_2 \rangle \vert}
  \vert \psi_1 \rangle + b 
  \frac{\langle \chi \vert \psi_1 \rangle}{\vert \langle \chi \vert \psi_1 \rangle \vert}
  \vert \psi_2 \rangle \right) \label{eq14}.
\end{equation}
The success probability in this case, $P_{3}=\frac{c_1 c_2}{c_1+c_2}N_{\psi}^2$.
Comparing the success probabilities resulting from these two protocols, we have,
\begin{eqnarray}
 r_p = \frac{P_2}{P_3} &=& \frac{c_1+c_2}{2(c_1\vert a \vert^2+c_2\vert b \vert^2)} \nonumber \\
 &=& \frac{r_c+1}{2(1+\vert b \vert^2 (r_c-1))},
\end{eqnarray}
where $r_c=\frac{c_2}{c_1} \in (0,\infty)$, $\vert a \vert^2, \vert b \vert^2 \in (0,1)$,
and $r_p \in (0, \infty)$. Same value of success probabilities ($P_2$ and $P_3$) 
result, in case the overlaps, $c_1=c_2$ or the superposition is obtained with
equal weights, i.e. $\vert a \vert^2=\vert b \vert^2$. Figure~\ref{plot} shows the 
variation $r_p$ vs $r_c$ at different values of $\vert b \vert^2$. It is interesting
to note that our two-qubit based protocol outperforms the three-qubit 
based protocol (in terms of success probabilities) 
in the range $ 0.5 < \vert b \vert^2 < 1$ (when $ 0 < r_c < 1$) 
and in the range $ 0 < \vert b \vert^2 < 0.5$ (when $ 1 < r_c < \infty$).
With reference to Table~\ref{table1}, experimental dataset numbered $7$ 
has $r_c=3$, $\vert b \vert^2=0.2$ and dataset numbered $8$ corresponds to $r_c=3$, $\vert b \vert^2=0.1$,
that correspond to $r_p>1$ as per Figure~\ref{plot}.
\subsection{Enhancement in success probability subject to prior information \label{special}}
In general, there is an interplay between the success probability with 
which the desired superposed state is obtained and the amount of prior 
information regarding constituent states. We impose certain constraints on 
the constituent states and observe its impact on the success probabilities.
Reconsidering the problem of superposition of two single-qubit states having 
fixed non-zero overlaps,
$\vert \langle  \chi \vert \psi_1 \rangle \vert^2=c_1$ and 
$\vert \langle  \chi \vert \psi_2 \rangle \vert^2=c_2$ with the referential
state $\vert \chi \rangle$~\cite{oszmaniec-prl-2016},
we have, $\vert \langle  \chi^{\bot} \vert \psi_1 \rangle \vert^2=c_1^{\bot}
=1-c_1$ and $\vert \langle  \chi^{\bot} \vert \psi_2 \rangle \vert^2
=c_2^{\bot}=1-c_2$, where $\langle \chi \vert \chi^{\bot} \rangle=0$.
In this case, we consider the action of the identity operator $U_1
=I \otimes I \otimes (\vert \chi \rangle \langle \chi \vert + 
\vert \chi^{\bot} \rangle \langle \chi^{\bot} \vert)$ (instead of 
$I \otimes I \otimes \vert \chi \rangle \langle \chi \vert$).
Using the overlaps of the input states with both $\vert \chi \rangle$ 
and $\vert \chi^{\bot} \rangle$, we observe an increase in the success 
probability (see Appendix~\ref{sec9}). Further, we implement the single-qubit unitary 
operator $ U_{\chi}$ ($U_{\chi^{\bot}}$) on the first qubit, if the 
third qubit is in state $\vert \chi \rangle$ ($\vert \chi^{\bot} 
\rangle$)(see Appendix~\ref{sec9} for details). The explicit forms of the operators are 
 \begin{displaymath} U_{\chi} = \frac{1}{N_1}
 \left(\begin{array}{cc} \frac{1}{\sqrt{c_1}} & \frac{1}{\sqrt{c_2}} 
 \\ \frac{1}{\sqrt{c_2}} & \frac{-1}{\sqrt{c_1}}
 \end{array}  \right); \hspace*{2mm}  
 U_{\chi^{\bot}} = \frac{1}{N_2}
 \left(\begin{array}{cc} \frac{1}{\sqrt{c_1^{\bot}}} & \frac{1}{\sqrt{c_2^{\bot}}} 
 \\ \frac{1}{\sqrt{c_2^{\bot}}} & \frac{-1}{\sqrt{c_1^{\bot}}}
 \end{array}  \right),  \end{displaymath}
where $N_1=\sqrt{(c_1+c_2)/c_1c_2}$, 
$N_2=\sqrt{(c_1^{\bot}+c_2^{\bot})/c_1^{\bot}c_2^{\bot}}$. In this formalism, 
we mainly study two types of constraints, both $\vert \psi_1 \rangle$ and 
$\vert \psi_2 \rangle$ lie in the $(i)$ same longitudinal plane, and 
$(ii)$ same transverse plane of the Bloch sphere, In case $(i)$, the 
desired superposed state is obtained with success probability,
 \begin{equation}
  P^{tot}=
  N_{\psi}^2 \left( \frac{c_1c_2}{c_1+c_2} + 
  \frac{c_1^{\bot}c_2^{\bot}}{c_1^{\bot}+c_2^{\bot}} \right)= 
  P_3 + N_{\psi}^2 \frac{c_1^{\bot}c_2^{\bot}}{c_1^{\bot}+c_2^{\bot}}  \label{eq11}
 \end{equation}
For $c_1=c_2^{\bot}$, the success probability, $P^{tot}=2P_3$, becomes 
double to that of the ordinary case. In case $(ii)$, we have $c_1=c_2
=c~\textrm{(say)}$, which implies $c_1^{\bot}=c_2^{\bot}=c^{\bot}~\textrm{(say)}$. 
Further, assuming both states occupy diametrically opposite positions  
on the respective spherical sections of the Bloch sphere, the total 
success probability obtained then is given by:
 \begin{equation}
  P^{tot}=N_{\psi}^2 \left( \frac{c}{2} + \frac{c^{\bot}}{2} \right) 
  = \frac{1}{2}N_{\psi}^2, \label{eq12}
 \end{equation}
 which is again greater than $P_3$. Further, if both states lie in 
 the equatorial plane, this pair of states becomes orthogonal, and 
 the success probability reaches $1/2$. Eqs.~\ref{eq11},~\ref{eq12} 
 give higher success probabilities (for certain $a,b$ values) as compared to 
 the $a,b$-dependent protocol discussed in Ref.~\cite{oszmaniec-prl-2016}.
 Recently, we came across a different approach~\cite{doosti-pra-2017},
 analyzing the superposition of arbitrary pair of orthogonal states.
\section{Conclusions \label{conclusion}}
We have experimentally created superposition of single-qubit states in
the defined framework, covering all possible aspects, i.e. (i) creation 
of various single-qubit states and obtaining their superposition, (ii) 
superposition with arbitrary weights, and (iii) superposition of 
single-qubit states in the presence of assumed overall phases. All the 
experimental results have been obtained with  fidelities over 0.97. 
This protocol has also been extended for the superposition of multiple 
states of a qudit. We have also discussed certain special cases where 
the desired superposed state is obtained with enhanced success probability.
\begin{acknowledgments}
  SD acknowledges the financial
support by The Institute of Mathematical Sciences Chennai India, 
Technische Universit\"{a}t Dortmund Germany,
and support by the International Collaborative Research Centre TRR 160 
``Coherent manipulation of interacting spin excitations in tailored semiconductors,"
funded by the Deutsche Forschungsgemeinschaft.
SD, GT, and SG would like to thank Somshubhro Bandyopadhyay, Manik Banik, 
Prathik Cherian J., Guruprasad Kar, Samir Kunkri, and Ramij Rahaman
for useful discussions.
\end{acknowledgments}
%

\appendix
\section{Encoding scheme \label{appA}}
\par
Let us discuss the case of superposition of $n$ number of pure states of a qudit. 
Considering a $d$-dimensional referential state $\vert \chi \rangle_{d}$,
whose overlap (magnitude) with each of the constituent state is known.
Therefore, assuming $\vert \langle \chi \vert \Psi_j \rangle_{d} \vert^2=c_j$,
where $j \in \{1,2,\dots,n \}$.
Let $a_1,\, a_2,\dots a_{n}$ be the 
desired weights for creating superposition of $d$-dimensional states
$\vert \Psi_1 \rangle_{d}$, $\vert \Psi_2 \rangle_{d}$, $\dots$ $\vert \Psi_{n} \rangle_{d}$ 
respectively. 
We begin with the initial state,
\begin{equation}
 \frac{1}{N}(a_1' \vert 0 \rangle_{n}+a_2' \vert 1 \rangle_{n}+\dots+a_{n}' \vert n-1 \rangle_{n})
 \otimes \vert \Psi_1 \rangle_{d} \otimes \dots 
 \otimes \vert \Psi_{n} \rangle_{d}, \label{eq21}
\end{equation}
where $N$ is the normalization factor, which is equal to 
$\sqrt{\sum_{j=1}^{n} a_j'^2}$.
This state belongs to a $n\times (d)^{n}$ dimensional Hilbert space,
where the primed coefficients are,
\begin{equation}
 a_k'=\frac{a_k}{\prod_{(j\neq k,j=1)}^{n} {\vert \langle \chi \vert \Psi_j \rangle_{d} \vert}}
 =\frac{a_k}{\sqrt{\prod_{(j\neq k,j=1)}^{n} {c_j}}}.
\end{equation}
This initial state is then 
made to undergo a series of controlled-swap operations, 
$\mathcal{C}\mathcal{S}_{2,3}^1~\mathcal{C}\mathcal{S}_{2,4}^1 \dots \mathcal{C}\mathcal{S}_{2,n}^1$
where state of first spin acts as control. 
In order to describe the action of this operation, let us
reconsider the set of bases vectors of the control spin, 
 ($\vert k \rangle_{n}$, $k\in \{0, 1, \dots, n-1 \}$) 
in $n$-dimensional Hilbert space,
whenever the first spin (qu$n$it) is in state $\vert k \rangle_{n}$, states of 
first qudit (second spin) and the $(k+1)^{th}$ qudit ($k+2^{th}$ spin) get swapped.
The resulting state is of the form,
\begin{eqnarray}
 \frac{1}{N}&& (a_1' \vert 0 \rangle_{n} 
 \otimes \vert \Psi_1 \rangle_{d} \otimes \vert \Psi_2 \rangle_{d} \otimes \dots 
 \otimes \vert \Psi_{n} \rangle_{d} \nonumber \\
 &&
  +a_2' \vert 1 \rangle_{n} 
 \otimes \vert \Psi_2 \rangle_{d} \otimes \vert \Psi_1 \rangle_{d} \otimes \dots 
 \otimes \vert \Psi_{n} \rangle_{d}  +\dots \nonumber \\
 &&
 +a_{n}' \vert n-1 \rangle_{n}
 \otimes \vert \Psi_{n} \rangle_{d} \otimes \vert \Psi_3 \rangle_{d} \otimes \dots 
 \otimes \vert \Psi_{1} \rangle_{d}). \nonumber \\ \label{eq22}
\end{eqnarray}
This is then acted upon by a set of projection operators constructed 
using the referential state $\vert \chi \rangle_{d}$. Operator performing 
$n-1$ number of projections on qudits numbered 2 to $n$ (or spins 
numbered 3 to $n+1$ in the 1-qu$n$it $\otimes$ n-qudit system) is given as,
$I_{n \times n} \otimes I_{d \times d}
\otimes \bigotimes_{k=2}^{n}(\vert \chi \rangle_{d} \langle \chi \vert_{d})_k$,
where $k$ represents the qudit number. This helps to remove the phases
that may be occurring with the constituent states ($\vert \Psi \rangle_{d}$'s).
The resulting state is given as,
\begin{eqnarray}
 \frac{1}{N} \sum_{k=1}^{n} \left( a_k \left( \prod_{(j\neq k, j=1)}^{n}
 {\frac{\langle \chi \vert \Psi_j \rangle_{d}}{\sqrt{c_j}}} \right) \vert k-1 \rangle_{n} \vert \Psi_{k} \rangle_{d} 
\right) 
 \bigotimes_{m=1}^{n-1} \vert \chi \rangle_{d} \nonumber \\ \label{eq15}
\end{eqnarray}
Tracing out states of qudits numbered 2 to $n$, we are
left with a $n \times d$-dimensional state. 
Also, shedding the overall phases, the state in Eq.~(\ref{eq15}) is written 
in a simple manner,
\begin{equation}
 \frac{1}{N} (a_1 \vert 0 \rangle_{n} \vert \Psi_1 \rangle_{d} + a_2 \vert 1 \rangle_{n} \vert \Psi_2 \rangle_{d} + 
 \dots + a_{n} \vert n-1 \rangle_{n} \vert \Psi_{n} \rangle_{d}), \label{eq9}
\end{equation}
where $N=\sqrt{\sum_{i=1}^{n} \vert a'_i \vert^2}$. In case of superposition of two 
qubits with weights $a_1=a$ and $a_2=b$, above equation is reduced to,
\begin{equation}
  \frac{1}{N} \left( a \frac{\langle \chi \vert \Psi_2 \rangle}{\vert \langle \chi \vert \Psi_2 \rangle \vert}
  \vert 0 \rangle \otimes \vert \Psi_1 \rangle
 + b \frac{\langle \chi \vert \Psi_1 \rangle}{\vert \langle \chi \vert \Psi_1 \rangle \vert} 
 \vert 1 \rangle \otimes \vert \Psi_2 \rangle \right), \label{eq12}
 \end{equation}
This is the two-qubit encoded state, which after Hadamard implementation
on first qubit, followed by a projection operator $|0\rangle \langle 0| \otimes I$
gives rise to the expected superposed state given as,
\begin{equation}
  \frac{1}{\sqrt{2}N} \left( a \frac{\langle \chi \vert \Psi_2 \rangle}{\vert \langle \chi \vert \Psi_2 \rangle \vert}
  \vert \Psi_1 \rangle
 + b \frac{\langle \chi \vert \Psi_1 \rangle}{\vert \langle \chi \vert \Psi_1 \rangle \vert} 
 \vert \Psi_2 \rangle \right), \label{eq12n}
 \end{equation}
The additional factor $\frac{1}{N}=\sqrt{\frac{c_1 c_2}{c_1\vert a \vert^2+c_2\vert b \vert^2}}$. Thus we reduce the 
existing three-qubit based protocol described in~\cite{oszmaniec-prl-2016} to the present 
two-qubit based protocol described in the main text.
It is to be noted that 
the state Eq.~(\ref{eq12}) has already taken care of the 
overall phases of states ($\vert \Psi_1 \rangle$ 
and $\vert \Psi_2 \rangle$). 
\section{Prior information and success probabilities\label{sec9}} 
There is an interplay between the amount of prior information
needed to superimpose a pair of partially known single-qubit 
pure states and the success probability with which the resultant
superposed state is obtained. In this section, we discuss 
the superposition protocol for pair of single-qubit pure 
sates under additional constraints that further leads to 
enhanced success probability. 
We re-consider the problem of superposition of two arbitrary single 
qubit states with known non-zero overlaps,
$\vert \langle  \chi \vert \psi_1 \rangle \vert^2=c_1$ and 
$\vert \langle  \chi \vert \psi_2 \rangle \vert^2=c_2$ with the referential
single-qubit state $\vert \chi \rangle$.
Thus one can obtain the overlaps of the constituent states 
with $\vert \chi^{\bot} \rangle$ (single-qubit state orthogonal 
to $\vert \chi \rangle$). 
We have, $\vert \langle  \chi^{\bot} \vert \psi_1 \rangle \vert^2=c_1^{\bot}=1-c_1$ and 
$\vert \langle  \chi^{\bot} \vert \psi_2 \rangle \vert^2=c_2^{\bot}=1-c_2$.
Let us begin with a three-qubit initial state, similar to the 
one given in Eq.~(\ref{eq21}),
\begin{equation}
 (a \vert 0 \rangle + b \vert 1 \rangle) \otimes \vert \psi_1 \rangle \otimes \vert \psi_2 \rangle.
 \label{eq31}
\end{equation}
 This state is then acted upon by the same three-qubit controlled-swap operation 
 as described in Appendix (A), such that the resulting state is,
 \begin{equation}
 a \vert 0 \rangle \otimes \vert \psi_1 \rangle \otimes \vert \psi_2 \rangle
 + b \vert 1 \rangle \otimes \vert \psi_2 \rangle \otimes \vert \psi_1 \rangle.
 \label{eq32}
\end{equation}
 Consider the action of the identity operator $U_1=I \otimes I \otimes 
 (\vert \chi \rangle \langle \chi \vert + \vert \chi^{\bot} \rangle \langle \chi^{\bot} \vert)$
 on the three-qubit state given in Eq.~(\ref{eq31}).
 The resultant state is given as,
 \begin{eqnarray}
   && \left[ a \langle \chi \vert \psi_2 \rangle \vert 0 \rangle \vert \psi_1 \rangle
 + b \langle \chi \vert \psi_1 \rangle \vert 1 \rangle \vert \psi_2 \rangle \right] \otimes \vert \chi \rangle \nonumber \\
   &&+ \left[ a \langle \chi^{\bot} \vert \psi_2 \rangle \vert 0 \rangle \vert \psi_1 \rangle
 + b \langle \chi^{\bot} \vert \psi_1 \rangle \vert 1 \rangle \vert \psi_2 \rangle \right] \otimes \vert \chi^{\bot} \rangle.  \nonumber \\
 \label{eq33}
 \end{eqnarray}
Another controlled unitary operation is implemented on the first qubit, where
state of third qubit acts as control. Subject to the state of the third qubit
($\vert \chi \rangle$ or $\vert \chi^{\bot} \rangle$),
the action of this controlled operation is described (on the first qubit) as,
\begin{eqnarray}
 U_{\vert \chi \rangle}\vert 0 \rangle &\rightarrow& \frac{1}{N_1}
 \left(\frac{1}{\sqrt{c_2}} \vert 0 \rangle + 
 \frac{1}{ \sqrt{c_1}} \vert 1 \rangle \right), \nonumber \\
  U_{\vert \chi \rangle}\vert 1 \rangle &\rightarrow& \frac{1}{N_1}
 \left(\frac{1}{\sqrt{c_1}} \vert 0 \rangle - 
 \frac{1}{\sqrt{c_2}} \vert 1 \rangle \right), \nonumber \\
  U_{\vert \chi^{\bot} \rangle}\vert 0 \rangle &\rightarrow& \frac{1}{N_2}
 \left(\frac{1}{\sqrt{c_2^{\bot}}} \vert 0 \rangle + 
 \frac{1}{\sqrt{c_1^{\bot}}} \vert 1 \rangle \right), \nonumber \\
  U_{\vert \chi^{\bot} \rangle} \vert 1 \rangle &\rightarrow& \frac{1}{N_2}
 \left(\frac{1}{\sqrt{c_1^{\bot}}} \vert 0 \rangle -
 \frac{1}{\sqrt{c_2^{\bot}}} \vert 1 \rangle \right), \\
 \label{eq34}
 \textrm{where,} \frac{1}{N_1}&=&\sqrt{\frac{c_1c_2}{c_1+c_2}} \,
\textrm{and} \, 
\frac{1}{N_2}=\sqrt{\frac{c_1^{\bot}c_2^{\bot}}{c_1^{\bot}+c_2^{\bot}}}. \nonumber
\end{eqnarray}
Eq.~(\ref{eq33}) thus leads to,
 \begin{eqnarray}
   && \frac{a}{N_1} \left( \frac{\langle \chi \vert \psi_2 \rangle}{\sqrt{c_2}} \vert 0 \rangle 
   + \frac{\langle \chi \vert \psi_2 \rangle}{\sqrt{c_1}} \vert 1 \rangle \right) \vert \psi_1 \rangle \otimes \vert \chi \rangle \nonumber \\
 &+& \frac{b}{N_1} \left(\frac{\langle \chi \vert \psi_1 \rangle}{\sqrt{c_1}} \vert 0 \rangle
 - \frac{\langle \chi \vert \psi_1 \rangle}{\sqrt{c_2}} \vert 1 \rangle
 \right)  \vert \psi_2 \rangle  \otimes \vert \chi \rangle \nonumber \\
    &+& \frac{a}{N_2} \left( \frac{\langle \chi^{\bot} \vert \psi_2 \rangle}{\sqrt{c_2^{\bot}}} \vert 0 \rangle 
   + \frac{\langle \chi^{\bot} \vert \psi_2 \rangle}{\sqrt{c_1^{\bot}}} \vert 1 \rangle \right) 
   \vert \psi_1 \rangle \otimes \vert \chi^{\bot} \rangle \nonumber \\
 &+& \frac{b}{N_2} \left(\frac{\langle \chi^{\bot} \vert \psi_1 \rangle}{\sqrt{c_1^{\bot}}} \vert 0 \rangle
 - \frac{\langle \chi^{\bot} \vert \psi_1 \rangle}{\sqrt{c_2^{\bot}}} \vert 1 \rangle
 \right)  \vert \psi_2 \rangle  \otimes \vert \chi^{\bot} \rangle.
 \nonumber \\ 
 \label{eq35}
 \end{eqnarray}
 Application of the projection operator, $\vert 0 \rangle \langle 0 \vert \otimes I_{2 \times 2} \otimes I_{2 \times 2}$
 then leads to,
  \begin{eqnarray}
   && \frac{1}{N_1} \left( a \frac{\langle \chi \vert \psi_2 \rangle}
   {\vert \langle \chi \vert \psi_2 \rangle \vert} \vert \psi_1 \rangle
   +  b \frac{\langle \chi \vert \psi_1 \rangle}
   {\vert \langle \chi \vert \psi_1 \rangle \vert} \vert \psi_2 \rangle \right) \otimes \vert \chi \rangle \nonumber \\
    &+& \frac{1}{N_2} \left( a \frac{\langle \chi^{\bot} \vert \psi_2 \rangle}
    {\vert \langle \chi^{\bot} \vert \psi_2 \rangle \vert} \vert \psi_1 \rangle 
   + b \frac{\langle \chi^{\bot} \vert \psi_1 \rangle}
   {\vert \langle \chi^{\bot} \vert \psi_1 \rangle \vert} \vert \psi_2 \rangle  \right) \otimes \vert \chi^{\bot} \rangle.
 \nonumber \\ \label{eq36}
 \end{eqnarray}
 Thus we obtain the weighted superpositions of single-qubit states 
 $\vert \psi_1 \rangle$ and $\vert \psi_2 \rangle$. If state of second qubit here is 
 $\vert \chi \rangle$, the superposed state,
 \begin{equation}
  \vert \Psi^{(1)} \rangle = \frac{ N_{\psi}^{(1)}}{N_1} \left( a \frac{\langle \chi \vert \psi_2 \rangle}
   {\vert \langle \chi \vert \psi_2 \rangle \vert} \vert \psi_1 \rangle
   +  b \frac{\langle \chi \vert \psi_1 \rangle}
   {\vert \langle \chi \vert \psi_1 \rangle \vert} \vert \psi_2 \rangle \right) \label{eq37}
 \end{equation}
  is obtained with a 
 success probability, $P^{(1)}= (N_{\psi}^{(1)})^2 \frac{c_1c_2}{c_1+c_2}$. While 
 corresponding to second-qubit state $\vert \chi^{\bot} \rangle$, the superposed state,
 \begin{equation}
  \vert \Psi^{(2)} \rangle = \frac{ N_{\psi}^{(2)}}{N_2} \left( a \frac{\langle \chi^{\bot} \vert \psi_2 \rangle}
    {\vert \langle \chi^{\bot} \vert \psi_2 \rangle \vert} \vert \psi_1 \rangle 
   + b \frac{\langle \chi^{\bot} \vert \psi_1 \rangle}
   {\vert \langle \chi^{\bot} \vert \psi_1 \rangle \vert} \vert \psi_2 \rangle  \right) \label{eq38}
 \end{equation}
 is resulted with a success probability, 
 $P^{(2)}= (N_{\psi}^{(2)})^2 \frac{c_1^{\bot}c_2^{\bot}}{c_1^{\bot}+c_2^{\bot}}$. 
 $N_{\psi}^{(1)}$ and $N_{\psi}^{(2)}$ are the normalization factors of the 
 first qubit state when states of the second qubit are $\vert \chi \rangle$
 and $\vert \chi^{\bot} \rangle$ respectively in Eq.~(\ref{eq36}).
 States given in Eqs.~(\ref{eq37}) and~(\ref{eq38}) are weighted superpositions 
 of the same constituent states $\vert \psi_1 \rangle$ and $\vert \psi_2 \rangle$.
 But they may be different because of their possibly different relative phases.
 The situation of our interest arises when $\vert \Psi^{(1)} \rangle$ varies 
 from $\vert \Psi^{(2)} \rangle$ only upto a global phase. 
 Following are few special cases discussing such scenarios.
  \begin{figure}
 \centering
 \includegraphics[scale=1,keepaspectratio=true]{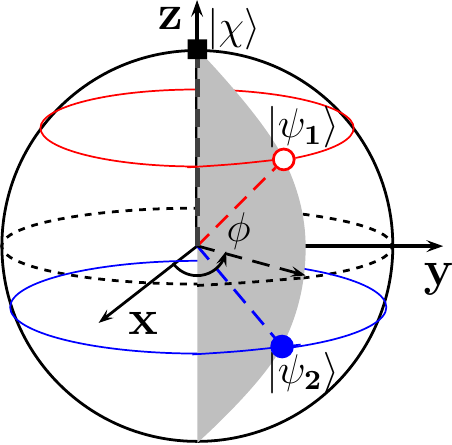}
 \caption{(Colour online) Bloch sphere representation of 
 $\vert \psi_1 \rangle$, $\vert \psi_2 \rangle$, and $\vert \chi \rangle$,
 marked with unfilled red circle, filled blue circle, 
 and filled black square respectively. 
 \label{sphere}} 
\label{sphere}
\end{figure}
 \subsubsection{Both states belong to same longitudinal plane on the Bloch sphere}
 Assume now that both $\vert \psi_1 \rangle$ and $\vert \psi_2 \rangle$ lie in the same 
 longitudinal plane on the Bloch sphere as shown in Fig.~\ref{sphere}.
 More explicitly, for $\frac{\langle \chi^{\bot} \vert \psi_j \rangle}
    {\vert \langle \chi^{\bot} \vert \psi_j \rangle \vert}=e^{\iota \phi}\frac{\langle \chi \vert \psi_j \rangle}
    {\vert \langle \chi \vert \psi_j \rangle \vert}$,
 Eq.~(\ref{eq36}) takes the form,
   \begin{eqnarray}
   && \left( a \frac{\langle \chi \vert \psi_2 \rangle}
   {\vert \langle \chi \vert \psi_2 \rangle \vert} \vert \psi_1 \rangle
   +  b \frac{\langle \chi \vert \psi_1 \rangle}
   {\vert \langle \chi \vert \psi_1 \rangle \vert} \vert \psi_2 \rangle \right) 
   \otimes \left( \frac{1}{N_1} \vert \chi \rangle + \frac{e^{\iota \phi}}{N_2} \vert \chi^{\bot} \rangle \right). \nonumber \\
\label{eq39}
 \end{eqnarray}
 Tracing out the second qubit, we obtain,
 \begin{eqnarray}
   && \sqrt{\frac{1}{N_1^2}+\frac{1}{N_2^2}} ~N_{\psi} \left( a \frac{\langle \chi \vert \psi_2 \rangle}
   {\vert \langle \chi \vert \psi_2 \rangle \vert} \vert \psi_1 \rangle
   +  b \frac{\langle \chi \vert \psi_1 \rangle}
   {\vert \langle \chi \vert \psi_1 \rangle \vert} \vert \psi_2 \rangle \right), \nonumber \\
\label{eq392}
 \end{eqnarray} 
 which is the desired superposed state. This superposed state is obtained with success probability,
 \begin{equation}
  P^{tot}=
  N_{\psi}^2 \left( \frac{c_1c_2}{c_1+c_2} + \frac{c_1^{\bot}c_2^{\bot}}{c_1^{\bot}+c_2^{\bot}} \right)= 
  P_3 + N_{\psi}^2 \frac{c_1^{\bot}c_2^{\bot}}{c_1^{\bot}+c_2^{\bot}}.  \label{seq11}
 \end{equation}
 This can as well be written as, $P^{\rm tot}=P + P^{\bot}$, where 
 $P=\left( \frac{N_{\psi}}{N_1} \right)^2$ and $P^{\bot}=\left( \frac{N_{\psi}}{N_2} \right)^2$. 
 Putting another constraint, $c_1=c_2^{\bot}$, 
 we obtain $N_1=N_2$ which gives rise to 
the desired superposed state with a success probability,
\begin{equation}
 P^{\rm tot}= 2 N_{\psi}^2 \frac{c_1c_2}{c_1+c_2} = 2 P.
\end{equation}
\subsubsection{Both states belong to same transverse plane on the Bloch sphere}
In this case, we have $c_1=c_2=c~\textrm{(say)}$, which implies
$c_1^{\bot}=c_2^{\bot}=c^{\bot}~\textrm{(say)}$. 
 Eq.~(\ref{eq35}) thus leads to,
  \begin{eqnarray}
   && \frac{1}{N}\vert 0 \rangle \left( a \frac{\langle \chi \vert \psi_2 \rangle}{\sqrt{c}} 
   \vert \psi_1 \rangle +b \frac{\langle \chi \vert \psi_1 \rangle}{\sqrt{c}}
  \vert \psi_2 \rangle \right) \otimes \vert \chi \rangle \nonumber \\
     &+& \frac{1}{N}\vert 1 \rangle \left( a \frac{\langle \chi \vert \psi_2 \rangle}{\sqrt{c}} 
   \vert \psi_1 \rangle -b \frac{\langle \chi \vert \psi_1 \rangle}{\sqrt{c}}
  \vert \psi_2 \rangle \right) \otimes \vert \chi \rangle \nonumber \\
     &+& \frac{1}{N}\vert 0 \rangle \left( a \frac{\langle \chi^{\bot} \vert \psi_2 \rangle}{\sqrt{c^{\bot}}} 
   \vert \psi_1 \rangle + b \frac{\langle \chi^{\bot} \vert \psi_1 \rangle}{\sqrt{c^{\bot}}}
  \vert \psi_2 \rangle \right) \otimes \vert \chi^{\bot} \rangle \nonumber \\
  &+& \frac{1}{N}\vert 1 \rangle \left( a \frac{\langle \chi^{\bot} \vert \psi_2 \rangle}{\sqrt{c^{\bot}}} 
   \vert \psi_1 \rangle - b \frac{\langle \chi^{\bot} \vert \psi_1 \rangle}{\sqrt{c^{\bot}}}
  \vert \psi_2 \rangle \right) \otimes \vert \chi^{\bot} \rangle. \nonumber \\
 \label{eq41}
 \end{eqnarray} 
Further, assuming both states occupy diametrically opposite positions  
on respective spheric sections of the
Bloch sphere, the azimuthal angles of the two states may be considered as
$\phi$ and $\pi+\phi$. Under the action of projection operator, $\vert 0 \rangle \langle 0 \vert
\otimes I \otimes \vert \chi \rangle \langle \chi \vert$ Eq.~(\ref{eq41}) gives rise to the 
desired superposed state,
  \begin{eqnarray}
    \frac{1}{N_1} \left( a \frac{\langle \chi \vert \psi_2 \rangle}{\sqrt{c}} 
   \vert \psi_1 \rangle +b \frac{\langle \chi \vert \psi_1 \rangle}{\sqrt{c}}
  \vert \psi_2 \rangle \right) 
 \end{eqnarray} 
 with a success probability, $P=(\frac{N_{\psi}}{N_1})^2$.
 Note that with the projection operator, $\vert 1 \rangle \langle 1 \vert
\otimes I \otimes \vert \chi^{\bot} \rangle \langle \chi^{\bot} \vert$ Eq.~(\ref{eq41}) gives rise to the 
desired superposed state,
  \begin{eqnarray}
    \frac{1}{N_2} \left( a \frac{\langle \chi^{\bot} \vert \psi_2 \rangle}{\sqrt{c^{\bot}}} 
   \vert \psi_1 \rangle +b \frac{\langle \chi^{\bot} \vert \psi_1 \rangle}{\sqrt{c^{\bot}}}
  \vert \psi_2 \rangle \right) 
 \end{eqnarray} 
 with a success probability, $P^{\bot}=(\frac{N_{\psi}}{N_2})^2$.
 The total success probability obtained in above two instances,
 \begin{equation}
  P^{\rm tot}=P+P^{\bot}=N_{\psi}^2 \left( \frac{1}{N_1^2} + \frac{1}{N_2^2} \right) = \frac{1}{2}N_{\psi}^2 \label{seq12}
 \end{equation}
 Further, if both states lie in the equatorial plane, 
 this pair of states becomes orthogonal, and the success probability reaches $1/2$.
 Eqs.~\ref{seq11},~\ref{seq12} give higher success probabilities (for certain $a,b$ values) as compared to 
 the $a,b$-dependent protocol discussed in the supplemental material of Ref.~\cite{oszmaniec-prl-2016}.
\end{document}